\documentclass[10pt]{article}
\usepackage[spanish,english]{babel}
\usepackage{amstext}
\usepackage{amssymb}
\usepackage{amsthm}
\usepackage{amsmath}

\newcommand{\mbb}[1]{\mathbb{#1}}

\newcommand{\mb}[1]{\boldsymbol{#1}}
\newcommand{\mc}[1]{\cal{#1}}
\newcommand{\ti}{\'{\i}}
\newcommand{\ket}[1]{\left | \, #1 \right \rangle}
\newcommand{\bra}[1]{\left \langle #1 \, \right |}

\newcommand{\fp}[1]{#1 \negthickspace :}

\begin{document}
\renewcommand{\thefootnote}{\fnsymbol{footnote}}

\selectlanguage{english}

\title{Some relations between quantum Turing machines and Turing machines}
\author{\textbf{Sicard, Andr\'es}\footnote{email address: asicard@sigma.eafit.edu.co} \textbf{ and V\'elez,
Mario}\footnote{email address: mvelez@sigma.eafit.edu.co}
\\
EAFIT University; Medell\ti n, Colombia, S.A.}
\maketitle

\renewcommand{\thefootnote}{\arabic{footnote}}

\begin{abstract}
For quantum Turing machines we present three elements: Its
components, its time evolution operator and its local transition
function. The components are related with the components of deterministic Turing
machines, the time evolution operator is related with the evolution of
reversible
Turing machines and the local transition function is related with
the transition function of 
probabilistic and reversible Turing machines.
\end{abstract}

\selectlanguage{spanish}
\begin{abstract}
  Para las m\'aquinas de Turing cu\'anticas se presentan tres elementos:
  Sus componentes, su operador de evoluci\'on temporal y su funci\'on
  de transici\'on local. Los componentes son relacionados con los
  componentes de las m\'aquinas de Turing determ\ti sticas, el operador de evoluci\'on
  temporal es relacionado con la evoluci\'on de las  m\'aquina de Turing reversibles y
  la funci\'on de transici\'on local es relacionada con la funci\'on
  de transici\'on de las m\'aquinas
  de Turing probabil\ti sticas y reversibles. 
\end{abstract}

\selectlanguage{english}

\section{Introduction}
First time, David Deutsch \cite{Deutsch-1985} described quantum TM
(Turing machine), he said the computational power of a quantum TM
and an classical (deterministic) one are the same for functions
from $\mbb{Z}$ to $\mbb{Z}$. However, due to inherent properties
of quantum TM, this equivalence is not immediate. Then it is
necessary to introduce some distinctions for Turing machines. In
particular, time evolution operator for quantum TM is unitary, it
means, evolution is reversible; this reason because it is
necessary to consider the quantum TM as a reversible one. On the
other hand, local transition function for a quantum TM represents
probability amplitude for evolution, in this case, it is necessary
to consider the quantum TM as a probabilistic one.

\section{The components: quantum TM and deterministic TM}
A TM has a finite set of states:

\begin{equation}\label{eq-10}
     Q = \{q_1, q_2, \dots, q_p\}.
\end{equation}
\
The states for an quantum TM are a finite set of observables:

\begin{equation}\label{eq-20}
  \mb{\hat{n}} = \{ \hat{n}_1,\hat{n}_2,\dots,\hat{n}_k \}.
\end{equation}
\
Every observable $\hat{n}_i \in \mb{\hat{n}}$ has spectrum
$\{0,1\}$ and every $q \in Q$ can take two values $\{current, non
\; current\}.$
\\
\\
The  TM works on a bi-infinity unidimensional tape. The machine
reads from the tape or writes over it some symbols, that belong to
a finite alphabet:

\begin{equation}\label{eq-30}
  \Sigma = \{0,1\}.
\end{equation}
\
Quantum TM represents the bi-infinity unidimensional tape by an
infinity set of observables:

\begin{equation}\label{eq-40}
  \mb{\hat{m}} = \{\hat{m}_i\}; \quad i \in \mbb{Z}.
\end{equation}
\
Every observable $\hat{m}_i \in \mb{\hat{m}}$ has spectrum
$\{0,1\}$ and every TM tape cell can take some values from the
alphabet $\Sigma = \{0,1\}$.
\\
\\
The TM has a read-write head. This head marks the position of the
machine on the tape. The quantum TM represents the read-write head
by an observable $\hat{x}$. The spectrum for the observable
$\hat{x}$ is $\mbb{Z}$ because of the existence of infinite cell
in TM tape.
\\
\\
The instantaneous description of a TM is formed by the current
state, the symbols on the tape and the read-write head's position.
The state of a quantum TM is a \cite{Deutsch-1985} ``\emph{unit
vector in the Hilbert space $\mc{H}$ spanned by the simultaneous
eigenvector:}

\begin{equation}\label{eq-50}
  \ket{\psi} = \ket{x,\mb{n},\mb{m}},
\end{equation}
\
\emph{of $\hat{x}$, $\mb{\hat{n}}$ and $\mb{\hat{m}}$, labelled by
the corresponding eigenvalues $x$, $\mb{n}$ and $\mb{n}$}''. The
states given by equation (\ref{eq-50}) are called the
computational bases states.

\section{The evolution: quantum TM and reversible TM}
Let $Q$ be states set and let $\Sigma$ be the alphabet for a TM.
The machine movements set (left, no motion, right) on the tape is
represented by:

\begin{equation}\label{eq-55}
    D = \{-1,0,1\}.
\end{equation}
\
\\
Evolution for machine is represented for a finite set of
instructions:

\begin{equation}
  \label{eq-60}
  q,s,s',d,q' \qquad \text{where} \quad q,q' \in Q; s,s' \in \Sigma; d \in
  D;
\end{equation}
\
\\
Instruction ``$q,s,s',d,q'$'' means: if the current state is $q$
and if the symbol on the cell marked by the read-write head is
$s$, the machine writes the symbol $s'$ in this cell, the machine
moves in to direction marked by $d$ and the machine goes to state
$q'$. Then, evolution for TM is a transition function $\delta$:

\begin{equation}
  \label{eq-65}
\fp{\delta} Q \times \Sigma \times \Sigma \times D \times Q \to
\{0,1\}, \quad \text{where},
\end{equation}

\begin{equation}
  \label{eq-70}
  \delta(q,s,s',d,q') =
  \begin{cases}
    1  & \text{iff $q,s,s',d,q'$ is an instruction for the TM},
    \\
    0 & \text{iff $q,s,s',d,q'$ is not an instruction for the
    TM}.
  \end{cases}
\end{equation}
\
\\
The TM is deterministic if and only if the transition function
$\delta$ satisfies:
\\
For any $(q,s) \in Q \times \Sigma$:
\begin{equation} \label{eq-80}
  \sum_{\substack{s' \in \Sigma \\ d \in D \\ q' \in Q}}\delta(q,s,s',d,q') \in
  \{0,1\}.
\end{equation}
\
\\
A deterministic TM is a reversible one if and only
if the transition function $\delta$ satisfies for any
$(q,s),(q',s') \in Q \times \Sigma$ with $(q,s) \neq (q',s')$:

\begin{equation}
  \label{eq-100}
  \sum_{\substack{s'' \in \Sigma \\ d \in D \\ q'' \in
  Q}}\delta(q,s,s'',d,q'') + \delta(q',s',s'',d,q'') \in \{0,1\}.
\end{equation}
\
\\
On the other hand, the evolution for a quantum TM during a single
computational step is \cite{Deutsch-1985, Ozawa-Nishimura-1999}:

\begin{equation}\label{eq-90}
\ket{\psi(t)} = U^t \ket{\psi(0)}, \quad t \in \mbb{Z}^+, \qquad
\text{where} \quad U^{\dagger}=U^{-1}.
\end{equation}
\
\\
$U$ is an unitary operator called time evolution operator. Because
$U$ is unitary, the evolution of a quantum TM is reversible.
Bennett proved that for any deterministic TM there is an
equivalent reversible TM \cite{Deutsch-1985}, in this way
irreversibility is not an essential feature of TM, while
reversibility is an essential feature of a quantum TM.

\section{The transition function: quantum TM and probabilistic TM}
Let $\tilde{\mbb{R}}$ be the computable real numbers set and
let$\widetilde{[0,1]}$ be the computable real numbers belong interval
$[0,1]$. If the transition function for a Turing Machine is:

\begin{equation}
  \label{eq-110}
\fp{\delta} Q \times \Sigma \times \Sigma \times D \times Q \to
\tilde{\mbb{R}},
\end{equation} and satisfies for any $(q,s) \in Q \times \Sigma$:

\begin{equation} \label{eq-115}
  \sum_{\substack{s' \in \Sigma \\ d \in D \\ q' \in Q}} \delta(q,s,s',d,q')
  \in \widetilde{[0,1]},
\end{equation}
\
\\
then the Turing machine is a probabilistic TM \cite{Gill-1977}.
The transition function $\delta$ of a probabilistic TM means that
if the current state is $q$ and if the symbol on the cell marked
by the read-write head is $s$, the probability of writing the
symbol $s'$ in this cell, of moving into the direction marked for
$d$ and of changing to current state $q'$ is given by the value of
function $\delta(q,s,s',d,q')$.
\\
\\
A Turing machine operates by finite means \cite{Turing-1936}. The
finite operation of a (deterministic, reversible, probabilistic)
TM is supported in finite process unit set and finite alphabet
(this implies a finite instructions set). On the other hand, a
quantum TM operates by finite means if and only if
\cite{Deutsch-1985}: ``\emph{only a finite subsystem is in motion
during any one step, and the motion depends only on the state of a
finite subsystem, and the rule that specifies that motion can be
given finitely in the mathematical sense}''.
\\
\\
To meet the requirements of finite operation, matrix elements of time
evolution (unitary) operator $U$ given by the equation (\ref{eq-90})
will have the form given by \cite{Deutsch-1985, Ozawa-Nishimura-1999}:
\\
\\
For any states $\ket{x,\mb{n},\mb{m}}$ and
$\ket{x',\mb{n}',\mb{m}'}$:

\begin{align}\label{eq-120}
& \bra{x',\mb{n}',\mb{m}'} U \ket{x,\mb{n},\mb{m}} = \notag \\ &
\quad \Bigr[\mb{\delta}_{x'}^{x+1}
\delta(\mb{n},m_{x},m_{x}',1,\mb{n}') + \mb{\delta}_{x'}^{x}
\delta(\mb{n},m_{x},m_{x}',0,\mb{n}')  + \notag \\ & \quad \;
\mb{\delta}_{x'}^{x-1} \delta(\mb{n},m_{x},m_{x}',-1,\mb{n}')
\Bigl] \prod_{y \neq x}\mb{\delta}_{m_{y}}^{m_{y}'}
\end{align}
\
\\
where $\mb{\delta}$ is Kronecker delta and $\delta$ is local
transition function for a quantum TM.
\\
\\
Let $\tilde{\mbb{C}}$ be the complex computable numbers set, let
$\mb{n} \in \mb{N} = \{0,1\}^k$ be, where $\mb{\hat{n}}$ is given
by equation (\ref{eq-20}); let $\mb{m} \in \mb{M} =
\{0,1\}^{\mbb{Z}}$ be, where $\mb{\hat{m}}$ is given by equation
(\ref{eq-40}) and let $m_x \in \Sigma$ be, where $\Sigma$ is given
by equation (\ref{eq-30}). The function $\delta$ is a function
\cite{Bersntein-Vazarini-1997, Ozawa-Nishimura-1999}:

\begin{equation}
  \label{eq-130}
\fp{\delta} \mb{N} \times \Sigma \times \Sigma \times D
\times \mb{N} \to \tilde{\mbb{C}}.
\end{equation}
\
If the state of a quantum TM is $\ket{x,\mb{n},\mb{m}}$ and $d=x'-x$,
the function $\delta(\mb{n},m_x,m_x',d,\mb{n}')$
represents probability amplitude for to evolve to state
$\ket{x',\mb{n}',\mb{m}'}$.
\\
\\
From the equation (\ref{eq-120}) it is possible determinate the
local transition function $\delta$ from time evolution
operator $U$. Conversely, is possible determinate the time evolution
operator $U$ from the local transition function $\delta$ by
\cite{Ozawa-Nishimura-1999}:

\begin{equation} \label{eq-140}
  U\ket{x,\mb{n},\mb{m}} = \sum_{\substack{m_x' \in \Sigma \\ d \in D
  \\ \mb{n}' \in \mb{N}}}
  \delta(\mb{n},m_x,m_x',d,\mb{n}')\ket{x+d,\mb{n}',\mb{m}'}.
\end{equation}
\
\\
Local transition $\delta$ should (indirectly) satisfy equation
(\ref{eq-90}). Operator $U$ is unitary if and only if $\delta$
satisfies the following conditions \cite{Ozawa-Nishimura-1999}:
\\
\\
\begin{enumerate}
\item For any $(\mb{n},m_x) \in \mb{N} \times \mb{M}$:

\begin{equation} \label{eq-150}
  \sum_{\substack{m_x' \in \Sigma \\ d \in D \\ \mb{n}' \in \mb{N}}} \mid
  \delta(\mb{n},m_x,m_x',d,\mb{n}') \mid ^2 = 1.
\end{equation}

\item For any $(\mb{n}, \mb{m}, x), (\mb{n}', \mb{m}', x') \in \mb{N} \times \mb{M} \times \mbb{Z}$ with
$(\mb{n}, \mb{m}, x) \neq (\mb{n}', \mb{m}', x')$:

\begin{equation} \label{eq-160}
  \sum_{\star}
  \delta(\mb{n}',m_{x'}',m_{x'}^{\diamond},d',\mb{p})^{*}\delta(\mb{n},m_x,m_x^{\diamond},d,\mb{p})
= 0.
\end{equation}
\
\\
where, the summation $\underset{\star}{\sum}$ is taken over all
$\mb{p} \in \mb{N}$; $\mb{m}^{\diamond} \in \mb{M}$; $d, d' \in D$
and $x \in \mbb{Z}$ such that $x + d = x' + d'$.
\end{enumerate}
\
\\
Equation (\ref{eq-150}) is quantum counterpart for equation
(\ref{eq-115}), under the relation between probability and
probability amplitude. Equation (\ref{eq-160}) is quantum
counterpart for equation (\ref{eq-100}), it means, the local
transition function $\delta$ should be reversible.

\section{Conclusion}
There are some ``classic'' models (deterministic, reversible,
probabilistic) equivalent (from computability's point of view) for
a Turing machine. Other model, the quantum Turing machine can be
compared to these models. From a ``physics'' perspective, it
means, from $U$ evolution operator's point of view, a quantum TM
can be seen as a reversible TM and from a ``mathematical''
perspective, it means, from the $\delta$ local transition
function's  point of view, a quantum TM can be seen how an
probabilistic and reversible TM.

\section{Acknowledgements}

The paper was financed by EAFIT University, under the research
project number $817407$.

\bibliographystyle{acm}
\bibliography{qtm}
\end{document}